# Plasmon-polaron of the topological metallic surface states

Alex Shvonski and Krzysztof Kempa

Department of Physics, Boston College, Chestnut Hill, Massachusetts 02467, USA

## Abstract

We report a plasmon-polaron mode of a 2D electron gas occupying surface states of a 3D topological crystal. This new, low-frequency, acoustic plasmon mode splits-off from the conventional spin-plasmon mode as a result of strong interactions of the surface electrons with bulk phonons. We show that, like in the case of the conventional spin-plasmon, the scattering of this mode is strongly suppressed in some regions of the phase space. This signature of the topological protection leads to an Umklapp-free mode dispersion at the Brillouin zone boundary. Such a plasmon polaron mode has indeed been recently observed in the topological metal $Be_2Se_3$.

Topological insulators (TI) are novel systems, with topologically protected metallic helical electronic surface states, characterized by suppressed back scattering, and Dirac-like linear dispersions at the center of the Brillouin zone. Like any normal (topologically trivial) insulator, TI has bulk and surface phonon modes [1], but at the surface (in contrast to the topologically trivial insulators) it also supports helical plasmons (spin-plasmons) [2,3]. Since the helical surface electronic states are topologically distinct from those in the bulk, their dynamics is affected by phonons, only through their contributions to an effective dielectric response. In this paper, we demonstrate that a new 2D collective electronic mode exists on the surface of TI, which emerges from such a polarization induced coupling between the surface electrons and bulk phonon modes. We also establish a correspondence between the effective dielectric function of this system and the polaron model proposed by Bozovic [4].

We employ the random phase approximation (RPA) to describe the surface collective electronic modes, and account for the phonon contributions through the Fröhlich electron-phonon-electron term, included in the effective, background dielectric function. We show that, in addition to the conventional spin-plasmon, there exists a second, low frequency, collective, polaron-like mode. By calculating explicitly the scattering of this mode with a periodic perturbing potential of the crystal lattice, we show that its backscattering is strongly suppressed, and that this suppression results from topological protection. Finally, we show that the recently observed collective mode on the surface of the strongly doped topological metal $Be_2Se_3$ has all signatures of such a spin-plasmon polaron [5].

The plasmon problem in a TI was considered [2,3,6] by assuming the Hamiltonian $H_0 = \hbar v_F (k_x \sigma_y - k_y \sigma_x)$, where $\sigma_x$ and $\sigma_y$ are Pauli matrices acting in the space of the electron spin projections, and the helical eigenfunction of the Hamiltonian is $e^{i\mathbf{k}\cdot\mathbf{r}} |f_\mathbf{k}\rangle / S$ (S is surface area), and the spinor part of it is

$$|f_\mathbf{k}\rangle = \frac{1}{\sqrt{2}} \begin{pmatrix} e^{-i\varphi_\mathbf{k}/2} \\ i e^{i\varphi_\mathbf{k}/2} \end{pmatrix} \tag{1}$$

where $\varphi_\mathbf{k}$ is the polar angle of the vector **k**, and the electron energies are $E_\mathbf{k} = v_F k$. The expectation value of the electron spin is therefore $\langle f_\mathbf{k} | \sigma | f_\mathbf{k} \rangle = \hat{\mathbf{z}} \times \mathbf{k}$ (where $\hat{\mathbf{z}}$ is the vector perpendicular to the 2D gas). Using the RPA, it was shown that the condition for existence of a the plasmon is

$$\varepsilon_{2D} = 1 - V_q^{2D} \Pi(q,\omega) = 0 \tag{2}$$

where $V_q^{2D} = 2\pi e^2 / q \varepsilon_{eff}$. For a single subband, the relevant situation here, we have

$$\Pi(q,\omega) = \frac{1}{S} \sum_\mathbf{k} |\langle f_{\mathbf{k+q}} | f_\mathbf{k} \rangle|^2 \frac{n_\mathbf{k} - n_{\mathbf{k+q}}}{\hbar\omega + E_\mathbf{k} - E_{\mathbf{k+q}} + i0^+} \tag{3}$$

where $n_\mathbf{k}$ is the Fermi – Dirac occupation number. Eq. (2) predicts a quasi-2D plasmon mode, the dispersion of which is $\omega \propto \sqrt{q}$, for small $q \ll k_F$, and $\omega \approx \beta q$, for larger $q$. This is the characteristic form of dispersion for all 2D gases (normal and topological).

The 2D electron gas resides on the surface of the topological metal (insulator with partially filled conduction band). Thus, the Coulomb interaction must be modified to include contributions from both electrons and phonons in the bulk of the topological insulator. The shielding of Coulomb electronic interactions is now controlled by an effective $\varepsilon_{eff}$, defined by the following expression (exact in RPA) [7,8]

$$V_{eff}(q,\omega) = \frac{V_q}{\bar{\varepsilon}} + \frac{\Omega_q |g_q/\bar{\varepsilon}|^2}{\omega^2 - \omega_q^2 + i\delta} = \frac{V_q}{\varepsilon_{eff}} \qquad (4)$$

where $V_q^{3D} = 4\pi e^2/q^2$, $\bar{\varepsilon} \approx [1+\varepsilon]/2$, $\varepsilon$ is the background dielectric constant due to electrons in the TI bulk, and the acoustic phonon dispersion is given (approximately) by

$$\omega_q \approx \alpha q \qquad (5)$$

The second term in the square bracket of Eq. (4) is the Fröhlich term.

It is straightforward to show that for $q \gg k_F$, Eq. (2) reduces to

$$\varepsilon_{2D} \approx 1 + \frac{r_s}{4}\left(\frac{\bar{\varepsilon}}{\varepsilon_{eff}}\right)\left(\frac{k_F}{q}\right)^2 = 0 \qquad (6)$$

where the ratio of the Coulomb interaction energy to the kinetic energy is $r_s = e^2/\bar{\varepsilon}\hbar v_F$.

Eq. (6) leads to $\varepsilon_{eff} = -\frac{\bar{\varepsilon} r_s}{4}\left(\frac{k_F}{q}\right)^2 \to 0^-$ (or $1/\varepsilon_{eff} \to -\infty$), which according to Eq. (4) can occur only for

$$\omega \approx \omega_q \approx \alpha q \qquad (7)$$

The analysis above shows that there exist two modes of the 2D electron gas on the surface of a TI. One is the conventional 2D spin plasmon, as discussed just below Eq.(3), with dispersion $\omega \propto \sqrt{q}$ for $q \ll k_F$, and $\omega \approx \beta q$ for larger $q$. A sketch of this mode (for large $q$) is shown in Fig. 1 (black-dashed line). The second, low frequency plasmon results from strong electron-phonon coupling, which electromagnetically renormalizes the 2D plasmon dispersion. Thus, this mode has a strong polaron character. This mode has linear dispersion for large q (Eq. 7), as shown in Fig. 1 (blue-dashed line). Clearly, the two collective electronic 2D modes are well separated in frequency for all *q*, and thus not coupled. The electronic susceptibility is the same for both modes, and given by Eq. (3).

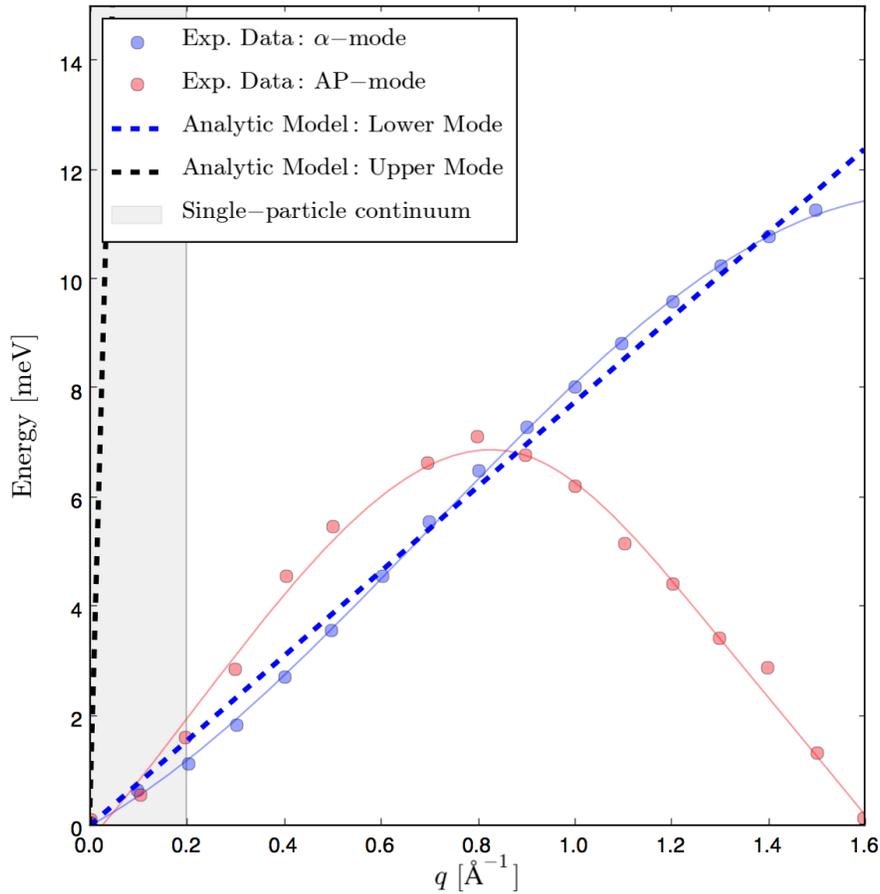

**Fig. 1**. Plasmonic modes of the 2D electron gas on the surface of a topological metal (parameters for $Be_2Se_3$). Lines represent calculated dispersions: upper mode $\omega \approx \beta q$ (black-dashed line) and $\omega \approx \omega_q \approx \alpha q$ (blue-dashed line). Blue circles represent experimental data for $Be_2Se_3$ in the topological, and the red circles in the normal states [5]. The shaded region represents the single particle continuum.

Fig. 1 also shows the experimental dispersions for metal $Be_2Se_3$ in the topological state and normal state (after Mn doping), obtained in Ref. [5], which are represented by blue and red circles, respectively. The corresponding single-particle excitation range, extending up to q ~ $2k_F$, is shown in Fig. 1 as a shaded region.

One of the most important discoveries in Ref. [5] was that the quasi-linear polaron mode dispersion surprisingly extends into the second BZ without an expected Umklapp at the M-point on the BZ edge. This is equivalent to an absence of a gap opening at this point. To understand this effect, consider the situation in which the 2D electron gas is subjected to a weak periodic potential, which explicitly represents the atomic lattice of the underlying TI. The plasmon problem can be written as an eigenvalue problem of a general electron-hole Hamiltonian [3,8-10], with plasmon eigenstate $|\mathbf{q}\rangle$ and corresponding eigenvalue being the plasma frequency. The projection of the eignestate, i.e., the momentum wavefunction, is given by

$$\langle \mathbf{k}|\mathbf{q}\rangle = A_q \frac{\langle f_{\mathbf{k+q}}|f_{\mathbf{k}}\rangle}{\omega_q + E_{\mathbf{k}} - E_{\mathbf{k+q}} + i0^+} = B_{kq}\langle f_{\mathbf{k+q}}|f_{\mathbf{k}}\rangle \qquad (8)$$

where $A_q$ is a normalizing factor. In this situation, we can apply a perturbation theory for the degenerate plasmon eigenstates at the *M*-point on the edge of the BZ: $|\mathbf{q}_M\rangle$ and $|\mathbf{q}_M - \mathbf{G}\rangle$, (see Fig. 2). Both states have the same, unperturbed frequency $\omega_{q=q_M}$. The perturbed frequency is

$$\tilde{\omega}_{q=q_M} = \omega_{q=q_M} \pm \Delta \qquad (9)$$

where a gap of the size $2\Delta$ opens at the edge of BZ for non-vanishing matrix element

$$\Delta = \left|\langle \mathbf{q}_M | V | \mathbf{q}_M - \mathbf{G} \rangle\right| \qquad (10)$$

Inserting an identity operator $I = \sum_{\mathbf{k}} |\mathbf{k}\rangle\langle\mathbf{k}|$ and Eq. (8) into Eq. (10), and assuming that $k \ll q_M$ and $k \ll |\mathbf{q}_M - \mathbf{G}|$ we get

$$\Delta = \left|\sum_{\mathbf{k}} \langle \mathbf{q}_M | \mathbf{k} \rangle V \langle \mathbf{k} | \mathbf{q}_M - \mathbf{G} \rangle\right| \approx \left|V\left(B_{0q_M}\right)^* B_{0|\mathbf{q}_M - \mathbf{G}|} \sum_{\mathbf{k}} \langle f_{\mathbf{k}+\mathbf{q}_M - \mathbf{G}} | f_{\mathbf{k}} \rangle \langle f_{\mathbf{k}} | f_{\mathbf{q}_M + \mathbf{k}} \rangle\right| \qquad (11)$$

It can be easily shown using Eq. (1) that

$$\sum_{\mathbf{k}} \langle f_{\mathbf{k}+\mathbf{q}_M - \mathbf{G}} | f_{\mathbf{k}} \rangle \langle f_{\mathbf{k}} | f_{\mathbf{q}_M + \mathbf{k}} \rangle = C \langle f_{\mathbf{q}_M - \mathbf{G}} | f_{\mathbf{q}_M} \rangle \qquad (12)$$

where $C$ is a constant.

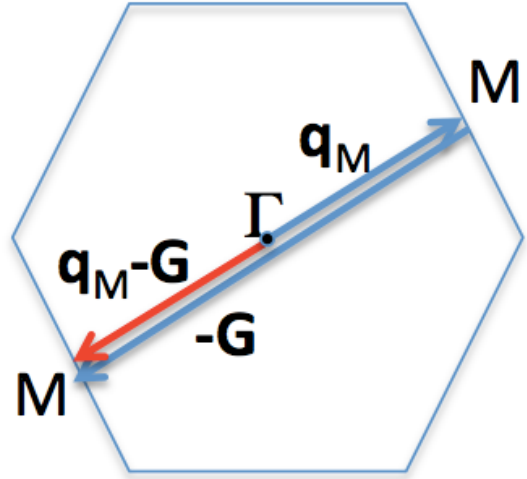

**Fig. 2**. Configuration of plasmonic vectors in the first BZ. $|\mathbf{q}_M - \mathbf{G}\rangle$ represents the backwards scattered plasmon-polaron state from the initial state $|\mathbf{q}_M\rangle$ at the M-point to the *M*-equivalent point on the other side of the BZ.

Since, according to Eq. (1),

$$\langle f_{\mathbf{q}_M-\mathbf{G}} | f_{\mathbf{q}_M} \rangle = \cos(\theta/2) \tag{13}$$

and $\theta$ is the angle between $\mathbf{q}_M$ and $\mathbf{q}_M - \mathbf{G}$. Since $\theta = \pi$, according to Fig. 2, the spinor inner product given by Eq. (13) vanishes, which leads to

$$\Delta \approx 0 \tag{14}$$

Note, that by crystal symmetry $\Delta' = |\langle \mathbf{q}_M | V | \mathbf{q}_M + \mathbf{G} \rangle|$ must vanish similarly, since the final state in this forward scattering $|\mathbf{q}_M + \mathbf{G}\rangle$ is also at the *M*-equivalent point. Thus, *both the backwards and forward scattering of the polaron-plsmon are suppressed*, and this important result is a direct consequence of the helical character of the electronic states.

This result can be confirmed by noticing that $\Delta$ given by Eq. (10) is simply related to the plasmon angular scattering form factor $\Phi(q,\theta)$, studied in detail in Ref. [3]

$$\Delta = |\langle \mathbf{q}_M | V | \mathbf{q}_M - \mathbf{G} \rangle| \propto \sqrt{\Phi(|\mathbf{q}_M|,\theta)} \tag{15}$$

$\Phi(q,\theta)$ can be numerically evaluated as a function of the scattering angle, and we have done this for parameters of the polaron mode in $Be_2Se_3$. Fig. 2 shows the corresponding polar plot. There is a dramatic suppression of scattering for both, forward ($\theta = 0$), and backwards scattering ($\theta = \pi$). Thus, according to Eq. (15), $\Delta \approx 0$, which is in agreement with Eq. (14), and demonstrates that the opening of the gap at the M-point is strongly suppressed due to suppressed backscattering of the plasmon polaron mode.

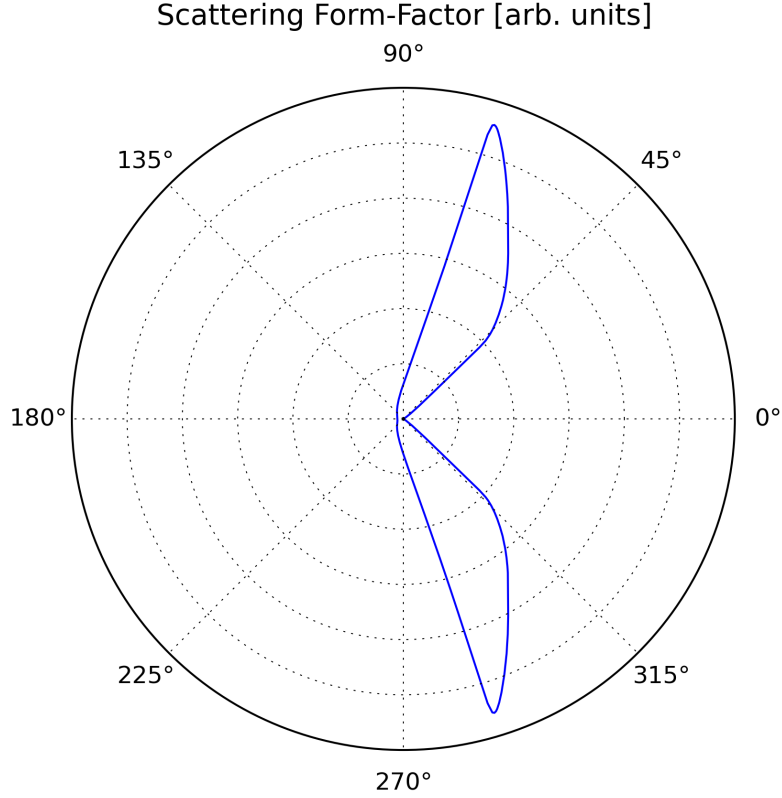

**Fig. 3**. Polar plot of the plasmon angular scattering form factor $\Phi(q,\theta)$. Parameters are $q = 4.44 p_F$, $\mu = 300\, meV$, $r_s = 0.09$, and $\varepsilon = 40$, where $\mu = p_F v_F$, as defined in [3]. The plasmon dispersion from Fig. 1 (blue dashed line) was used.

Our theory has a direct correspondence to the simple coupled-oscillator model proposed by Bozovic [4]. The Bozovic polaron is a classical, coupled state of an electron (with mass $m$) and a neutral, polarized particle (mass $M$), with the dielectric function given by

$$\varepsilon_D = 1 - \frac{\omega_p^2\left(\omega^2 - \Omega^2\right)}{\omega^2\left(\omega^2 - \omega_0^2\right)} \qquad (16)$$

where $\omega_0^2 = \kappa(m+M)/mM$ and $\Omega^2 = \kappa/M$, and $\omega_0^2 > \Omega^2$. To see the correspondence with our nonlocal theory, we write a hydrodynamic analog of Eq. (2), which contains the main dynamics of the polaron mode, as

$$\varepsilon_{2D} = 1 - \frac{\bar{\omega}_p^2 q}{\varepsilon_{eff}\left(\omega^2 - \beta^2 q^2\right)} = 0 \qquad (17)$$

which, with the help of Eq. (4) can be reduced to the form of Eq. (16). This shows, that the polaron mode in our theory is a quantum analog of the classical Bosovic polaron, and results from the strong coupling to the polarizable dielectric background of the TI lattice.

In conclusion, we have shown that a 3D topological crystal supports a low-frequency plasmon-polaron mode of its surface 2D electron gas. As a result of strong interaction of the 2D electrons with bulk phonons, the plasmon mode existence condition yields two collective modes: the conventional, gaples spin-plasmon, and the low-frequency acoustic, plasmon-polaron. Due to large phase space separation of these modes (except for very small momenta), there is little interaction between the modes. Since the 2D electron gas is topologically distinct from the bulk, its dynamics is affected by phonons, but in contrast to topologically trivial situations, only through their contributions to an effective dielectric response of the environment as seen by the surface electrons. We show that, not unlike in the case of the conventional spin-plasmon, the scattering of this mode is strongly suppressed in some regions of the phase space, which leads to an Umklapp-free mode dispersion at the Brillouin zone boundary. The low-lying plasmon mode, with an Umklapp-free behavior has indeed been recently observed in the topological crystal $Be_2Se_3$.